# Implementation of Secure Quantum Protocol using Multiple Photons for Communication

Sayonnha Mandal[1], Gregory Macdonald[1], Mayssaa El Rifai[1], Nikhil Punekar[1], Farnaz Zamani[1],
Yuhua Chen[2], Subhash Kak[3], Pramode K. Verma[1][†], Robert C Huck[1], James Sluss[1]
[1] Telecommunications Engineering Program, School of Electrical and Computer Engineering,
University of Oklahoma-Tulsa, Tulsa, OK 74135, USA, [†] Email: pverma@ou.edu
[2] Dept. of Electrical and Computer Engineering, University of Houston, Houston, TX 77204, USA
[3] Department of Computer Science, Oklahoma State University, Stillwater, OK 74078, USA

*Abstract*—The paper presents the implementation of a quantum cryptography protocol for secure communication between servers in the cloud. As computing power increases, classical cryptography and key management schemes based on computational complexity become increasingly susceptible to brute force and cryptanalytic attacks. Current implementations of quantum cryptography are based on the BB84 protocol, which is susceptible to siphoning attacks on the multiple photons emitted by practical laser sources. The three-stage protocol, whose implementation is described in this paper, is a departure from conventional practice and it obviates some of the known vulnerabilities of the current implementations of quantum cryptography. This paper presents an implementation of the three-stage quantum communication protocol in free-space. To the best of the authors' knowledge, this is the first implementation of a quantum protocol where multiple photons can be used for secure communication.

## I. INTRODUCTION

As cloud services gain popularity, it is inevitable that more and more government and business entities outsource their computational needs to the cloud. However, securing information within the cloud between the servers is of paramount importance. One can use classical cryptography means to secure the information transferred between the servers. However, classical implementations of cryptography are only computationally secure. As computing power increases, classical cryptography and key management schemes based on computational complexity become increasingly susceptible to brute force and cryptanalytic attacks. On the other hand, the security of quantum cryptography is based on the inherent uncertainty in quantum phenomena. It is the only known means of providing secure communication [1] regardless of computational power. However, current implementations of quantum cryptography are based on the BB'84 protocol [2], which is susceptible to siphoning attacks on the multiple photons emitted by practical laser sources. This makes BB'84-based quantum cryptography protocol unsuitable for the cloud computing environment.

In this paper, we propose a secure quantum cryptography implementation that allows multiple photons for communications between the servers in the cloud. The proposed secure multi-photon quantum cryptography implementation is based on the mathematical formation of the three-stage protocol discovered by one of the co-authors [3]. The proposed quantum cryptography implementation is a departure from conventional practice and it obviates some of the known vulnerabilities of the current implementations of quantum cryptography. To the best of the authors' knowledge, this is the first implementation of a quantum protocol where multiple photons can be used for secure communication. We present the background of quantum cryptography as follows.

Quantum cryptography is the only known mechanism for providing secure communication [4]. Key distribution is one of the most important issues in cryptography. Quantum key distribution (QKD) techniques under development today are aspiring to play a center-stage role in distributing keys for encryption and decryption. In QKD systems, which are commercially available, quantum states are used for encoding information.

The BB84 protocol is the most adopted QKD protocol in applications and experiments for quantum key distribution. At present, two companies - MagiQ and IDQ - have successfully developed commercial and R&D products for quantum key distribution. Unfortunately, BB84 was recently hacked [5] although patches were soon put in place by the manufacturers.

Most of the quantum key distribution protocols available so far utilize the polarization states of photons to realize the key distribution. Quantum physics states that not only is it generally impossible to gain any precise information about the unknown polarization state of a single photon but also that, once the polarization of the photon is measured, its polarization is irreversibly altered. Therefore, had we not known the exact polarization of the photon before it was measured, a measurement, will, in general, not reveal that exact polarization. Due to the no-cloning theorem, an unknown quantum state of a single photon cannot be copied.

The proposed multi-photon quantum cryptography implementation is based on the use of unitary commutative transformations known only to either Alice or Bob. This protocol gives an effective alternative to the BB84 protocol. Another contrasting factor is that the transmitted information bit in the proposed implementation can be in any arbitrary state of polarization and does not require being in one of the four states in the BB84 scheme [3].

From the point of view of practical implementations, the advantage is that information exchange is not restricted to the existence of only a single photon in a time slot. Even if the laser pulse produces multiple photons, as long as all the photons are in the same phase, the transformation and their complex conjugate transformation will have the same effect on them. Therefore, the proposed implementation is not subject to the beam-splitting attack.

Actual implementations of quantum cryptography protocols generally lag the theoretical constructs by a significant amount of time. The realization of the BB84 protocol beyond the research lab potentially took a couple of decades. In this paper, we implement the multi-photon quantum protocol in an experimental setup utilizing passive optical hardware components over free space. As implementation of a quantum cryptography protocol that is not susceptible to siphoning attacks, it makes the protocol well suited to the cloud environment.

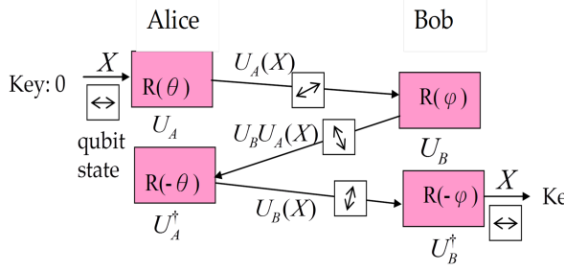

Figure 1. Schematic diagram of the three-stage protocol

The paper is organized as follows. In Section II, the three-stage protocol is described in detail and the specific transformation that is utilized in the proposed implementation scheme. In Section III, the actual implementation of the three-stage protocol using optical hardware components is discussed along with the schematic and detailed description of its operation. We conclude our work in Section IV.

## II. THE THREE-STAGE PROTOCOL

### A. Principle of Operation

We now discuss the method of operation of the three-stage protocol in terms of transferring state $X$ from Alice to Bob. The state $X$ is one of two orthogonal states, such as $|0>$ and $|1>$, or $\frac{1}{\sqrt{2}}(|0\rangle+|1\rangle)$ $\frac{1}{\sqrt{2}}(|0>+|1>)$ and $\frac{1}{\sqrt{2}}(|0\rangle-|1\rangle)$ $\frac{1}{\sqrt{2}}(|0>-|1>)$ or $\alpha|0\rangle+\beta|1\rangle$ $\alpha|0>+\beta|1>$ and $\beta|0\rangle+\alpha|1\rangle$. The orthogonal states of $X$ represent 0 and 1 by prior mutual agreement of the parties and this is the data or the cryptographic key being transmitted over the public channel.

Alice and Bob apply secret transformations $U_A$ and $U_B$ which are commutative, i.e., $U_A U_B = U_B U_A$.

An example of this would be $U_A = R(\theta)$ and $U_B = R(\varphi)$, each of which is the rotation operator:

$$R(\theta) = \begin{bmatrix} \cos\theta & -\sin\theta \\ \sin\theta & \cos\theta \end{bmatrix}$$

The summarized steps shown in Fig. 1 are described as follows:

Step 1: Alice applies a unitary transformation $U_A$ on quantum information $X$ and sends the qubits to Bob.

Step 2: Bob applies $U_B$ on the received qubits $U_A$, thereby giving $U_B U_A(X)$ and sends it back to Alice. $U_A$ and $U_B$ should be commutative transformations.

Step 3: Alice applies $U_A\dagger$ (transpose of the complex conjugate of $U_A$) on the received qubits to get $U_A^\dagger U_B U_A(X) = U_B(X)$ and sends it back to Bob.

Step 4: Then Bob applies $U_B\dagger$ on $U_B(X)$ to get the information $X$.

### B. Presence of Eve

We now consider the presence of an intruder on the communication channel [6]. Let us say the eavesdropper Eve intercepts the message transmitted between Alice and Bob. If Eve tries to differentiate between two non-orthogonal states, it is not possible to achieve information gain without collapsing the state of at least one of them. This is clear from considering |ψ> and |φ> to be the non-orthogonal quantum states that Eve is trying to know about. If these states interact with a standard state |v>,

|ψ>|v> ⟶ |ψ>|v>
|φ>|v> ⟶ |φ>|v′>

Eve would want |v> and |v′> to be different, in order to know the identity of the state. However inner products are preserved under unitary transformations and $<v|v′><ψ|φ> = <v|v><ψ|φ>$ or, $<v|v′> = <v|v> = 1$.

Therefore, |v> and |v′> must be identical and Eve will need to disturb one of the two states in order to acquire any information. Quantum key distribution is thereby effective because of the no-cloning theorem which makes this kind of attack ineffective.

However, the above analysis assumes that single photons are used in the communication, as required by the BB'84 quantum cryptography protocol and its variants. Unfortunately, practical implementations of BB'84 use attenuated lasers and it is possible that multiple photons are emitted. Therefore, practical implementations of BB'84 protocol are not secure with the presence of Eve. In contrast, the proposed implementation of the three-stage protocol allows multiple photons to be used in the secure communication, even with the presence of Eve. This is of critical importance in a practical cloud environment.

*C. Transformations*

The original three-stage protocol described in [3] provides the mathematical foundation of the quantum protocol implemented in this paper. As mentioned earlier, the realization of any quantum cryptography protocol usually lags its theoretical counterpart due to significant challenges faced in dealing with practical quantum systems. In this section, we discuss the implementation aspects of the protocol and practical realization of the rotation operators, which are crucial to providing secure data transfers. The section also highlights the use of transformations that apply on multiple qubits simultaneously.

One possible implementation is to apply Pauli transformations. They are convenient to use, and entail less precision requirements. The only condition for applying any new transformation in the operation of the three-stage protocol is that the transformations should map into the |0> and |1> states with equal probability so that the requirement for cryptographic security remains intact. The simplest group consist of the basic single-qubit operators *I, X, Y, Z*:

$$I = \begin{pmatrix} 1 & 0 \\ 0 & 1 \end{pmatrix}, X = \begin{pmatrix} 0 & 1 \\ 1 & 0 \end{pmatrix}, Y = \begin{pmatrix} 0 & -i \\ i & 0 \end{pmatrix}$$
$$\text{and } Z = \begin{pmatrix} 1 & 0 \\ 0 & -1 \end{pmatrix}.$$

Alice and Bob pick any of the four operators to operate upon *S*. A secret *S*= |0> will become a |1> with 50% probability upon the use of one of these four transformations, establishing security of this procedure.

Another possible transformation is described as follows. Let Alice and Bob choose *K* and *L* from the set of transformations, which are the "do-nothing" and the Hadamard transformations, which also form a group:

$$K = \begin{pmatrix} 1 & 0 \\ 0 & 1 \end{pmatrix} \text{ and } L = \frac{1}{\sqrt{2}} \begin{pmatrix} 1 & 1 \\ 1 & -1 \end{pmatrix}$$

Once again, this provides security since there is a fifty percent chance that the initial state has been put into a superposition.

A two-qubit system can be generalized by considering transformations on several qubits at the same time, which only requires that the transformations remain commutative. This is accomplished as follows. For example, Alice and Bob may use the transformations $U_A$ and $U_B$ given below:

$$U_A = \begin{pmatrix} 1000 \\ 0100 \\ 0001 \\ 0010 \end{pmatrix} \text{ and } U_B = \begin{pmatrix} 0100 \\ 1000 \\ 0010 \\ 0001 \end{pmatrix}$$

In another example, for 2-qubit transformations, the protocol might be for Alice and Bob to either perform the DFT or do nothing at all. For the two-qubit case, the transformation is:

$$U_A = U_B = \frac{1}{2} \begin{bmatrix} 1 & 1 & 1 & 1 \\ 1 & i & -1 & -i \\ 1 & -1 & 1 & -1 \\ 1 & -i & -1 & i \end{bmatrix}$$

Another option is to use the four unitary 4×4 matrices that map into quaternions [10]:

$$\begin{bmatrix} 0 & 1 & 0 & 0 \\ -1 & 0 & 0 & 0 \\ 0 & 0 & 0 & 1 \\ 0 & 0 & -1 & 0 \end{bmatrix} \begin{bmatrix} 0 & 0 & 0 & -1 \\ 0 & 0 & -1 & 0 \\ 0 & 1 & 0 & 0 \\ 1 & 0 & 0 & 0 \end{bmatrix}$$
$$\begin{bmatrix} 0 & 0 & -1 & 0 \\ 0 & 0 & 0 & 1 \\ 1 & 0 & 0 & 0 \\ 0 & -1 & 0 & 0 \end{bmatrix} \begin{bmatrix} 1 & 0 & 0 & 0 \\ 0 & 1 & 0 & 0 \\ 0 & 0 & 1 & 0 \\ 0 & 0 & 0 & 1 \end{bmatrix}$$

As mentioned before, they commute due to the fact that global phase may be ignored. These matrices are tensor products of the Pauli matrices of (*X*) and, therefore, these transformations represent the same operations as in (*X*) when considered in pairs.

We have discussed various transformations that can be utilized to operate the three-stage quantum key distribution protocol. In the rest of this section, we describe the rotation operation which is used in our laboratory implementation. Note that in rotation operators, one must also account for the rotation caused by the turnaround of the qubit at each re-transmission, and take into account the difficulties with maintaining fidelity in the presence of noise.

The unitary transformations used in the three-stage protocol can be achieved through rotations of the polarization state of photons. The amount of these rotations can be changed for each encoded bit providing a higher level of robustness for this protocol. The three stage protocol implemented with encoding qubits with different polarization states can be described as follows:

*Step 1*: Alice encodes information bits with two specific polarization states; for example, she encodes bit "0" with 0° and bit "1" with 90°, and then she transforms the polarization of the encoded bits by an arbitrary polarization rotation $\theta_a$ and then sends them to Bob.

*Step 2*: Bob transforms the polarization of the received information bits with an arbitrarily polarization angle $\theta_b$ then sends them back to Alice.

*Step 3*: Alice transforms the polarization of the received information bits by - $\theta_a$ and sends them back to Bob.

*Step 4*: Bob then applies the transformation – $\theta_b$ and obtains the original qubits and measures them with his detectors.

The amount of polarization rotation that both sides (Alice and Bob) select to give to the information bits is an arbitrary and independent value which varies from 0 to 360 degrees. The schematic diagram of the three-stage protocol, using polarization rotation as the unitary transformation, is shown in Fig. 2.

### III. IMPLEMENTATION

In our experimental set-up, we use two representations of Alice and Bob each to reduce the complexity. It is considered that they operate in a single line and the three stage configuration is obtained by utilizing plane reflecting mirrors in their path. Moreover, it is further assumed that these two Alices and two Bobs are mutually trusted parties and authenticated to each other.

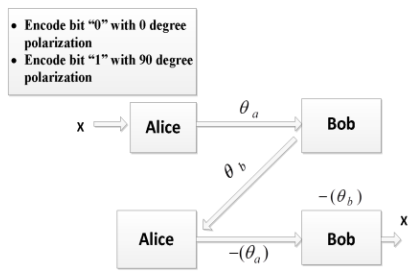

Figure 2. Schematic diagram of the three-stage protocol with using polarization rotation as the unitary transformation

### A. Hardware Specific Unitary Transformations Utilizing Muller Matrices

The unitary transformations of the three-stage protocol in the proposed implementation are applied using half-wave plates. A half-wave plate retards by $(-2\theta)$, where $\theta$ is the angle between the fast axis and the direction of polarization of the incident light. This type of wave plate changes the polarization direction of linear polarized light. This phase retardation property of half-wave plates is utilized in our experimental setup. The Muller matrix of a half-wave plate is given as follows:

$$M = \begin{bmatrix} 1 & 0 & 0 & 0 \\ 0 & \cos(4m) & \sin(4m) & 0 \\ 0 & \sin(4m) & -\cos(4m) & 0 \\ 0 & 0 & 0 & -1 \end{bmatrix}$$

where $m$ is the angle of rotation of the half-wave plate.

In our implementation, Alice uses two half-wave plates, at $x$ and $-x$ ($x$ is the angle of rotation of her half-wave plate) with Muller Matrices $M1$ and $M2$, respectively, and Bob uses two half-wave plates, at $y$ and $-y$ ($y$ being the angle of rotation of his half-wave plate) with Muller Matrices $M3$ and $M4$ respectively.

Assuming $x=30$ and $y=40$ and our incident light is horizontally polarized denoted by $S=$ [1 1 0 0]. The output O at Bob's detectors will be:

$$O = S*M_1*M_2*M_3*M_4 = S \quad \ldots\ldots.(1)$$

For $x$ and $y \in [0, 360]$, Eqn. (1) is valid.

### B. Principle of Operation with Schematic

The entire experimental procedure of the hardware based three-stage protocol over free space optics (FSO) shown in Fig. 3 is described in detail as follows:

**Encoding Stage**

1) In the hardware configuration, the light source is a HeNe linearly polarized Laser with a wavelength of 632.8 nm and power higher than 0.8 mW and 500:1 extinction ratio. The beam of light coming out of the laser is linearly polarized.
2) The laser light emanating from the same is incident on a 50/50 beam splitter. After passing through the beam splitter, the beam divides into two beams, each equal in intensity. The two beams travel through two different paths.
3) The beam of light traveling on the upper path (parallel to the path before the split) passes through a mechanical beam shutter (operating at a sustained maximum rate of 25 Hz with a minimum on time of 10 ms). The beam on the lower path passes through another identical shutter.

The two shutters (shown in Figure 3) are controlled through software by LabView programs. The shutters are programmed in such a way that the shutter in the upper path opens only when the bit it receives is a"1" and remains closed when the bit is a "0"; whereas, the other shutter (in the path orthogonal to the path of the source of light) opens at a "0" and remains closed at a "1". At the onset of the experiment, it is assumed that the two beam shutters are in the closed position and the APT motors are homed to their default 0 degree location. (The APT motors control the angle of rotation of the half-wave plates belonging to Alice).

In the upper path of the schematic, the beam coming out of the shutter passes through a polarizing filter (in the range of 400-700 nm) aligned at 90 degrees, thereby ensuring that the photons coming out of this polarizer have a 90 degree polarization.

4) In the other path, the light beam similarly travels through a polarizing filter, which in this case, is aligned at an angle of 0 degrees and thus converts the polarization of the photons passing out from it having a polarization angle of 0 degrees.

Thus, the beam of light traveling through these paths essentially gets encoded according to the bit stream of the key or the message. The polarizing filters perform the operation of bit encoding in this implementation, encoding bit "0" with 0 degree polarization and bit "1" with 90 degree polarization, respectively [7].

**Rotation Transformation Stage**

1) The two light beams polarized in the previous steps are combined with the help of plane mirrors angled at 45 degrees in the paths and finally made to converge into a single beam utilizing a combiner, which is essentially a reverse beam splitter.
2) The final beam of light, containing photons polarized either in the 90 or 0 degree angles, is then passed progressively through four half-wave plates (for a wavelength of 632.8 nm) oriented at different angles.
3) These four wave plates represent Alice and Bob, two wave plates for Alice and two for Bob, according to the configuration of the three-stage protocol, in which the information is sent from Alice to Bob, back to Alice and then finally to Bob again.

For our implementation, two half-wave plates are mounted on mechanical rotators (with maximum rotation velocity of 25 deg/sec and a continuous range of 360 degrees) controlled by software using LabView programming. For the purpose of our experiment, these mechanical rotator mounted half-wave plates are designated as Alice, whereas the half-wave plates mounted on the fixed mounts are designated as Bob.

4) The half-wave plates are set at particular angles so

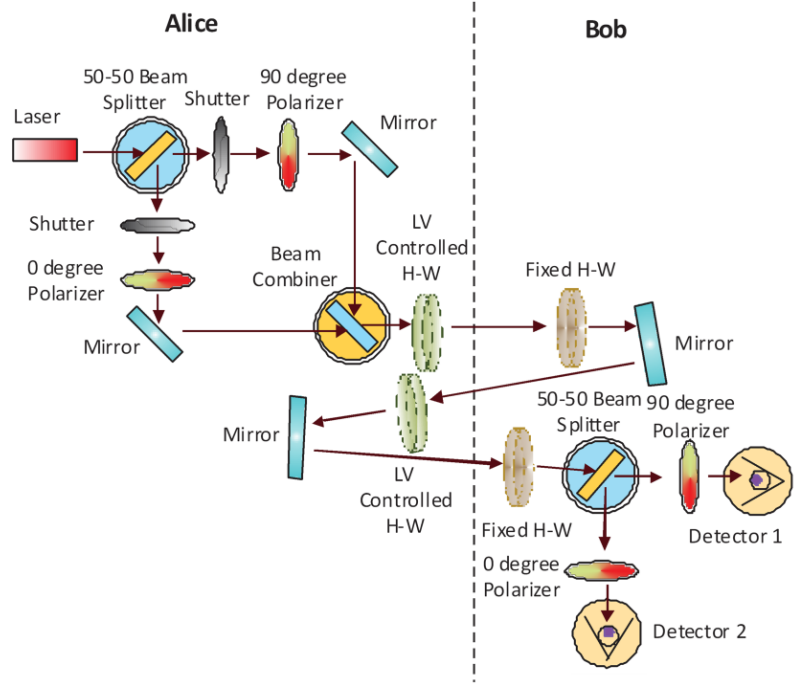

Figure 3: Schematic of the experimental set-up of the three-stage protocol over free space optics

that the photons passing through them have their polarizations changed by those specific angles. This configuration allows Alice to apply her rotation transformation and send bits to Bob.

5) Bob then applies his own rotation to the received bits and sends them back to Alice. The two fixed polarizers of Bob are set at angles that are inverse of each other from a polarization perspective.
6) Alice then finally inverts her angle of polarization and sends the bits to Bob, who then obtains the information by inverting his original rotation.
7) Alice then finally inverts her angle of polarization and sends the bits to Bob, who then obtains the information by inverting his original rotation.

**Decoding Stage**

1) In the experimental setup, the beam of light travels through the motorized half-wave plate (Alice 1) through the fixed half-wave plate (Bob 1) and then gets reflected by the mirror.
2) After passing through the motorized half-wave plate (Alice 2), it is reflected by another mirror (Mirror 4) to reach the fixed half-wave plate of Bob 2.
3) The light beam again passes through a beam splitter which divides the light into two beams where the photons are detected by a detector as a "1" or a "0", after passing through the 90 and 0 degree polarizers, respectively.

*C. LabView Results*

As mentioned earlier, LabView 10.0.1 is used to control the operation of the mechanical beam shutters and the motorized rotators for the wave plates. In the LabView program, the message to be transmitted is converted into bits and the shutters are opened according to bits "0" and "1" as described earlier in the paper. The angles of the wave plates are changed every eight bits, by rotating the half-wave plates by a random angle, using the APT motors.

After processing each block of eight bits, a character is generated and in this way the transmitted message sent from Alice's end over the free space is received at Bob's end.

IV. CONCLUSION

This is the first paper that presents an implementation of a quantum protocol which is not susceptible to siphoning attacks. It offers many advantages over the BB84 protocol for the cloud environment. More specifically, we have presented an implementation using rotation of polarization angles of the optical beam as the unitary transformations that Alice and Bob use in order to exchange information. The current implementation can be applied for exchanging keys but with higher data rates each data bit can be made secure. We expect the proposed implementation of multi-photon quantum cryptography protocol to have applications [8][9][10] for securing communications between servers in the cloud.

ACKNOWLEGEMENT

This research is supported in part by the National Science Foundation (NSF) under Grants 1117148, 1117179, and 1117068.